# Cationic Ordering and Microstructural Effects in the Ferromagnetic Perovskite $La_{0.5}Ba_{0.5}CoO_3$: Impact upon Magnetotransport Properties

*Eeva-Leena Rautama[1,2], Philippe Boullay[1], Asish K. Kundu[1], Vincent Caignaert[1], Valérie Pralong[1],*

*Maarit Karppinen[2], Bernard Raveau[1]*

[1]Laboratoire CRISMAT, UMR 6508 ENSICAEN/CNRS, 14050 Caen Cedex 4, France

[2]Laboratory of Inorganic and Analytical Chemistry, Helsinki University of Technology, P.O. Box 6100, FI-02015 TKK, Finland

**ABSTRACT**

The synthesis and structural study of the stoichiometric perovskite $La_{0.5}Ba_{0.5}CoO_3$ have allowed three forms to be isolated. Besides the disordered $La_{0.5}Ba_{0.5}CoO_3$ and the perfectly ordered layered $LaBaCo_2O_6$, a third form called nanoscale-ordered $LaBaCo_2O_6$, is obtained. As evidenced by transmission electron microscopy investigations, the latter consists of 112-type 90° oriented domains fitted into each other at a nanometer scale which induce large strains and consequently local atomic scale lattice distortions. These three ferromagnetic perovskites exhibit practically the same $T_C \cong 174$-179 K, but differently from the other phases, the nanoscale-ordered $LaBaCo_2O_6$ is a hard ferromagnet, with $H_C \approx 4.2$ kOe, due to the strains which may pin domain walls, preventing the reversal of the spins in a magnetic field. The magnetotransport properties of these phases show that all of them exhibit a maximum intrinsic magnetoresistance, close to 6-7 % around $T_C$ under ±70 kOe but that the ordered phase exhibits a much higher tunnelling magnetoresistance effect at low temperature of about 15 % against 4 % due to the grain boundary effects.

## INTRODUCTION

The fascinating magnetic and transport properties of manganites[1,2] and cobaltites[3-11] with the perovskite structure have been the object of numerous studies for the last ten years. Among these strongly correlated electron oxides, the stoichiometric perovskites with the generic formulation *Ln*Ba*M*$_2$O$_6$, where *Ln*=lanthanoide and *M*=Mn or Co, are of particular interest since they exhibit order-disorder phenomena between the Ln$^{3+}$ and Ba$^{2+}$ cations, which seem to significantly influence their magnetic and transport properties. For instance, the cubic disordered perovskite La$_{0.5}$Ba$_{0.5}$MnO$_3$ is ferromagnetic with a $T_C$ of 270 K, whereas the tetragonal 112-type ordered perovskite LaBaMn$_2$O$_6$, characterized by the 1:1 ordered stacking of LaO and BaO layers is also ferromagnetic, but with a much higher $T_C$ of 335 K[12]. Similarly, the cobaltites exhibit a disordered cubic La$_{0.5}$Ba$_{0.5}$CoO$_3$ form and a tetragonal 112-type layered LaBaCo$_2$O$_6$ form (see Figure 1) which are both ferromagnetic[13,14]. However, the influence of the cationic ordering upon the $T_C$ seems to be the reverse since, according to Nakajima et al.[14], the disordered perovskite exhibits a higher $T_C$ (190 K) than the ordered phase (175 K). More importantly, there is a discrepancy between the transport data of the disordered perovskite La$_{0.5}$Ba$_{0.5}$CoO$_3$ reported by the two groups[13,14]. Fauth et al.[13] observe for this phase a semi-metallic behavior with a metal-metal transition at $T_C$, whereas Nakajima et al.[14] observe a metallic behavior down to 140 K, with an abrupt increase in resistivity below this temperature. Recently, we synthesized a new form of this cobaltite, called nanoscale-ordered perovskite, which consists of 90°-ordered domains of the layered 112-type LaBaCo$_2$O$_6$[15,16]. The latter was shown to be also ferromagnetic but with a much higher coercivity than those of the two other forms which was attributed to a spin locking effect[16].

Bearing in mind the difficulty to control the ordering of the Ba$^{2+}$ and La$^{3+}$ cations in this structure, together with the oxygen stoichiometry, we report herein on the complex crystal chemistry of the cobaltites LaBaCo$_2$O$_6$ and La$_{0.5}$Ba$_{0.5}$CoO$_3$, and on its impact on the magnetotransport properties of these oxides. We show that three forms – disordered La$_{0.5}$Ba$_{0.5}$CoO$_3$, ordered LaBaCo$_2$O$_6$ and nanoscale-ordered LaBaCo$_2$O$_6$ – can be synthesized, using either soft chemistry or solid state reaction methods and controlling the oxygen partial pressure during the synthesis. The detailed cationic ordering of the La$^{3+}$ and $Ba^{2+}$ cations of these phases is studied, combining X-ray powder diffraction and high-resolution electron microscopy techniques emphasizing the peculiar behavior of the nanoscale-ordered perovskite $LaBaCo_2O_6$. Besides the spin locking effect recently discovered[16], the magnetotransport properties of the three forms are investigated, showing magnetoresistance ratios in the range of 6 to 15%. The relationships between the structure or nanostructure and the physical properties are discussed.

**EXPERIMENTAL TECHNIQUES OF CHARACTERIZATION**

The X-ray powder diffraction (XRPD) patterns were collected with a Philips X'pert Pro Diffractometer ($CuK_\alpha$ radiation) and refined according to the Rietveld method using the program JANA2006[17].

Composition analysis was carried out by energy dispersive spectroscopy (EDS) analysis using a Philips XL30 scanning electron microscope (SEM) equipped with an OXFORD analyzer which confirmed correct metal ratios within the detection limit of the instrument.

Oxygen contents of the samples were determined by iodometric titrations. About 50 mg of the sample powder was dissolved in acetic buffer solution under Ar atmosphere and the released iodine was then titrated with 0.1 M $Na_2S_2O_3$ solution in the presence of an indicator. Four to five parallel analyses were carried out for each sample with a reproducibility better than ±0.01 for the total oxygen content.

For the electron microscopy observations, the samples were crushed in an agate mortar in *n*-butanol and a drop of each suspension was deposited on a carbon-coated copper grid. Transmission electron microsopy (TEM) and high-resolution electron microscopy (HREM) studies were carried out with a JEOL 2010F electron microscope.

A Quantum Design physical properties measurements system (PPMS) was used to investigate the magnetic and electrical properties of the samples. Small pieces of rectangular bars were taken for the measurements. The field dependent magnetization, M(*H*), was recorded at different temperatures in an applied field of ±50 kOe. The electrical resistivity, ρ(*T*), and magnetoresistance, MR, measurements were carried out by a standard four probe method in the temperature range of 10-400 K with an applied field of ±70 kOe. The electrodes on the sample were prepared by ultrasonic deposition method using indium metal.

**CHEMICAL SYNTHESIS**

The cationic ordering in the LnBaCo$_2$O$_6$ perovskites strongly relates to the size difference between the Ln$^{3+}$ and Ba$^{2+}$ cations, similarly to what was observed for the 112-type manganites[12]. In the present LaBaCo$_2$O$_6$ case, the size difference between La$^{3+}$ and Ba$^{2+}$ is small, so that the ordering of these cations in the form of alternating layers is rather difficult to achieve. Here our strategy was to control the order-disorder phenomena in this system by means of two synthesis parameters, temperature and oxygen partial pressure.

In order to favour the ordering of the La$^{3+}$ and Ba$^{2+}$ cations, the synthesis temperature must be as low as possible, and consequently a soft-chemistry synthesis route should be used since it allows a high reactivity at low temperature. However, this condition is not sufficient alone to achieve a perfect ordering of these cations. The formation of the La$^{3+}$ layers seems in fact to be favoured by the intermediate creation of ordered oxygen vacancies, leading then to the 112-type layered oxides LnBaCo$_2$O$_{5.5}$, built up of layers of CoO$_5$ pyramids between which the smaller La$^{3+}$ cations can be interleaved. For this reason, the synthesis of the ordered phase was carried out in a reducive atmosphere, using high purity Ar gas, followed by annealing in an oxygen atmosphere at low temperature. Here, the ordered perovskite LaBaCo$_2$O$_6$ was successfully obtained through a sol-gel synthesis route. Stoichiometric amounts of La(NO$_3$)$_3$·4H$_2$O, Ba(NO$_3$)$_2$ and Co(NO$_3$)$_2$·6H$_2$O were first dissolved in water with the aid of stirring and heating (60°C). After obtaining a clear solution, citric acid (~5 times of molar amount) was added and the mixture was slowly evaporated at 120°C, until the gel formation was complete. Before weighing the nitrates, their actual water-of-crystallization contents were verified by thermogravimetric (TG) analysis (not shown here). The amorphous product was decomposed at 600°C for overnight and then calcined at 900°C for 12 h in air. After regrinding, the fine powder was pressed into pellets and sintered at 1150°C for 48 h in a flow of high purity Ar gas (5 ppm O$_2$) with heating and cooling rates of 2°C/min. Then to reach the "O$_6$" stoichiometry, the as-synthesized compound was oxygenated at 350°C under ~130 bar of oxygen for 12 h, using slow heating and cooling rates (1°C/min). The iodometric titration gave an oxygen content of 5.99 for the fully-oxygenated sample. Thus, this ordered perovskite can be formulated as $LaBaCo_2O_6$.

For the preparation of the disordered perovskite $La_{0.5}Ba_{0.5}CoO_3$, a solid state synthesis route was carried out similarly to the procedure used by Fauth *et al.*[13]. Stoichiometric mixtures of oxides $La_2O_3$ and $Co_3O_4$ and carbonate $BaCO_3$ were intimately ground and heated in air at 1200°C for 12 h and cooled rapidly down to room temperature. Then, the as-synthesized sample was annealed exactly under the same conditions as the ordered phase (350°C, $P_{O_2} \approx 130$ bar). Oxygen content of this disordered sample was determined from iodometric titration at 2.99, leading within the limits of experimental error to the formulation $La_{0.5}Ba_{0.5}CoO_3$.

Thus, the combination of the nature of precursors (soft chemistry or solid state reaction) and oxygen partial pressure is the key for the control of the cationic ordering in these materials. This is illustrated by changing the above experimental conditions of synthesis. For instance, using the first step of the sol-gel method described above, but heating the mixtures in air under the conditions described for the solid state reaction method, we were unable to obtain the disordered perovskite as a part of the $La^{3+}$ and $Ba^{2+}$ cations was always ordered at a short range. Likewise, using the sol-gel method described above for the synthesis of the ordered $LaBaCo_2O_6$ phase, but just heating under an oxygen flow instead of high purity Ar gas led to a complete disordering of the $La^{3+}$ and $Ba^{2+}$ cations. Based on these observations, we investigated the effect of a small variation in the oxygen pressure on the degree of cation ordering. Using the sol-gel method described above for the synthesis of the fully-ordered perovskite $LaBaCo_2O_6$, keeping all the same conditions of synthesis and sintering but just replacing the high purity Ar flow (5 ppm $O_2$) by a normal Ar flow (10 ppm $O_2$), we were able to obtain a new form that we call nanoscale-ordered perovskite whose microstructure and properties will be described below. The oxygen content, determined from iodometric titration as 6.01, allows us to give also the formula $LaBaCo_2O_6$ for the nanoscale-ordered perovskite.

**X-RAY DIFFRACTION STUDY**

The XRPD patterns of the three perovskite forms (Figure 2) show their excellent crystallization. The disordered phase La$_{0.5}$Ba$_{0.5}$CoO$_3$, (Figure 2a) as well as the ordered phase LaBaCo$_2$O$_6$ (Figure 2c) exhibit sharp peaks, and the patterns could be refined considering the structural data from the literature with the cubic *Pm-3m*[13] and tetragonal *P4/mmm*[14] space groups, respectively, the latter corresponding to a doubling of the cell parameter along the *c* axis related to the 1:1 ordering of the LaO/BaO layers (Figure 1). More importantly, the XRPD pattern of the nanoscale-ordered LaBaCo$_2$O$_6$ (Figure 2b) is very similar to that observed for the disordered phase (Figure 2a) and for this reason it was indexed and refined as a simple perovskite using the same space group *Pm-3m* and the same structural parameters (Table 1). However, enlargement of the patterns clearly shows an *hkl*-dependent peak broadening (inset of Figure 2b) compared to the disordered La$_{0.5}$Ba$_{0.5}$CoO$_3$ phase (inset of Figure 2a) that can be accounted for by considering an anisotropic strain broadening along the <100>$_p$ directions. Such a feature is closely related to the unique microstructure of this phase, as will be later shown on the basis of HREM observations. The refinements clearly show that the three forms exhibit practically the same cell volume per cobalt in agreement with the fact that the oxygen content is essentially the same for the three compounds. This result is slightly different from that obtained by Nakajima *et al.*[14], who observed a smaller cell volume for the ordered phase (58.66 Å$^3$/Co) compared to the disordered one (58.77 Å$^3$). Note also that the La/Ba ordering involves a slight deformation of the perovskite sublattice with a dilatation of the $a_p$ parameter within the LaO/BaO layers and a compression along the LaO/BaO layers stacking direction.

**ELECTRON MICROSCOPY INVESTIGATIONS**

The TEM investigations performed on the disordered perovskite La$_{0.5}$Ba$_{0.5}$CoO$_3$ and on the ordered layered perovskite LaBaCo$_2$O$_6$ confirm their structures. The selected area electron diffraction (SAED) patterns of the first one (inset of Figure 3a) and the corresponding HREM images (Figure 3a) are indeed characteristic of a classical cubic perovskite (*Pm-3m*) with $a \approx a_p \approx 3.9$Å. For the second one, the reconstruction of the reciprocal space from the SAED patterns leads to a tetragonal cell, with $a \approx a_p$ and $c \approx 2a_p$, compatible with the space group *P4/mmm*. The doubling of one cell parameter with respect to the simple perovskite cell is clearly observed on the HREM image displayed Figure 3b and on the corresponding SAED [100] zone axes patterns (ZAP) (inset of Figure 3b).

The SAED patterns of the nanoscale-ordered perovskite $LaBaCo_2O_6$ are more complex as illustrated in Figure 4a, for one of the $<100>_p$ ZAP typically observed for this compound, where only the strongest spots can be indexed on the basis of a simple perovskite cell. Besides, one indeed observes three sets of additional spots that can be indexed considering the perovskite cell, as $0\ k/2\ l$ and $0\ k\ l/2$, for two first sets, and as $0\ k/2\ l/2$, for a third set of weaker intensities. As indicated previously[15,16], these two sets of extra spots can be related to the existence, within the sample of two types of domains (denoted I and II) having the $2a_p$ supercell characteristic of the 112-type La/Ba ordering (see inset in the Figure 3b) and oriented at 90° with respect to each other. Tilting around the common $[001]^*_I/[010]^*_{II}$ direction (Figure 4a), one can evidence (Figure 4b) the existence of a fourth set of spots of weaker intensity, which correspond to another 112-type domain (III) oriented at 90° from the two others. The 3D reconstruction of the reciprocal space reveals that the structure of this compound is made up of 112-type ordered domains randomly distributed at a fine scale along the three equivalent $<100>_p$ directions of the perovskite subcell, leading to the "average" cubic simple perovskite observed in the XRPD patterns. Such a microstructural feature has already been observed in oxygen-deficient iron double perovskite[18,19] and referred to as a "microdomain texture"[20] schematically represented in Figure 4d. Coming back to the weaker set of spots ($0\ k/2\ l/2$), observed in the $<100>_p$ ZAP (Figure 4a), these reflections cannot be indexed simply considering a 112-type domain textured sample but could eventually be attributed to multiple scattering of the diffracted beams due to dynamical effects hardly avoidable in electron diffraction. Nonetheless, several experimental evidences (see for instance Figure 4c) indicate that the intensity observed at $0\ k/2\ l/2$ positions in the $<100>_p$ ZAP cannot only be attributed to multiple scattering but one should consider the existence of a secondary phase (denoted X). While this phase is likely a superstructure of the perovskite, it is difficult to know exactly which one, considering the 3D domain texture of the sample and the extra reflections can be indexed using various supercells.

Beside the SAED investigation which gave us a better view of the microstructure of the present perovskites, bright field (BF) images may be of great interest to obtain information about the size of the 112-type domains, taking into consideration the possibility of twinning. In the case of the disordered cubic simple perovskite no twinning is observed, as expected from the symmetry. In contrast, in the long-range ordered 112-perovskite $LaBaCo_2O_6$ twinning is clearly evidenced (Figure 5a). For the pseudo-cubic nanoscale-ordered perovskite, the BF image (Figure 5b) clearly attests that we are dealing here with a microstructural feature that goes beyond a simple twinned sample at a micrometer scale. Comparing the two images, the domains are not visible at this scale for the 3D domain textured sample. Instead, a specific mottled contrast consisting of curved and interpenetrated dark segments is observed, most probably related to the existence of strain fields. In order to image the domains, one has to go in higher magnification as illustrated in Figure 6. The BF image in Figure 6b is obtained by using an objective aperture whose size is chosen to exclude reflections from the perovskite subcell (Figure 6a). In the enlarged area zones, the $2a_p$ periodicity can be evidenced, with nanometer sized domains oriented at 90° (one domain type is dominant in this area). In the Bragg filtered image presented in Figure 6e where the contribution from the transmitted beam has been removed (Figure 6d), the parts of the image contributing to the formation of the spots at 0 k/2 l/2 positions are revealed (see arrows). The secondary phase appears here to be scarce, disseminated randomly and of smaller size than the 112-type domains. It also allows us to evidence various defects in the form of dislocations, crystallographic shears as well as lattice distortions from one domain to the other (compressed or dilated zones) which must be related with the mottled contrast observed in BF images.

From the BF image presented in Figure 6, taken in the vicinity of a $<100>_p$ zone axis orientation, it is not clear whether the 112-type domains are embedded in a disordered $(La,Ba)CoO_3$ perovskite matrix or not. The area without the $2a_p$ periodicity in Figure 6c can be attributed to zones having either a disordered $(La,Ba)CoO_3$ perovskite structure or a 112-type domain observed along the [001]* direction. This ambiguity is removed by examining the crystals along different crystallographic orientations (not shown here) and confirms that the 112-type domains are numerous and fitted into each other together with the secondary minority phase mentioned above. As shown in the HREM images in Figure 7a, taken along the <100>$_p$* direction (see the related SAED patterns Figure 4a), the 112-type domains are observed everywhere with a size ranging from 5nm to 10nm. At the first sight, only two types of 90° oriented domains seem to coexist (the ones with the in-plane 2$a_p$ periodicity). In the thicker part of the crystal, the 3D domain texture makes difficult to evidence the third 112-type domain with the out-of-plane 2$a_p$ periodicity. In Figure 7c, the Fourier transforms of the four zones indicated in the enlarged area in Figure 7b show how the three 90° oriented 112-type domains are fitted into each other in order. They form a projected 2D domain texture having (100)$_p$ planes as domain boundaries that would ideally lead to the 3D domain texture schematized in Figure 4d. This representation is largely oversimplified and actually domains can also be delimited by {110}$_p$ planes, as illustrated in the HREM image Figure 8, and more generally by a combination of both {100}$_p$ and {110}$_p$ planes at a local scale.

This study clearly demonstrates that beside the disordered La$_{0.5}$Ba$_{0.5}$CoO$_3$ and the 112-type long range ordered LaBaCo$_2$O$_6$ perovskites, the nanoscale-ordered LaBaCo$_2$O$_6$ perovskite must be considered as a third form, built up of 112-type nanometer-sized domains fitted into each other and 90° oriented. This specific 3D domain texture induces, in comparison to the two other forms, large strains in the material. There is no doubt that in such a strain-based material, the atomic-scale lattice distortions are coupled to the magnetic and electronic degrees of freedom. The latter will develop atomic elastic deformations and atomic relaxations in the vicinity of the ferromagnetic domain walls as described by Ahn *et al.*[21], in relation to the high coercive field observed in this material.

**MAGNETIZATION AND MAGNETOTRANSPORT PROPERTIES**

As previously demonstrated[14,16], in contrast to the perovskite manganites LaBaMn$_2$O$_6$ and La$_{0.5}$Ba$_{0.5}$MnO$_3$[12], the cationic order-disorder phenomena in these perovskite cobaltites do not affect their paramagnetic-to-ferromagnetic transition temperature, $T_C$. Our preliminary temperature dependent magnetization measurements also indeed show the $T_C$ of 177 K, 174 K and 179 K (calculated from the minimum position of the d$M_{FC}$/dT *versus* temperature plot) for the disordered La$_{0.5}$Ba$_{0.5}$CoO$_3$, ordered LaBaCo$_2$O$_6$ and nanoscale-ordered LaBaCo$_2$O$_6$, respectively. Moreover, other magnetic behaviors such as the field and the temperature dependent magnetization throughout the temperature range are similar to that of the reported behavior[13-14]. More importantly, the M(H) behavior of the nanoscale-ordered phase is different from the two other phases at low temperature (Figure 9). The coercive field, $H_C$, for the disordered $La_{0.5}Ba_{0.5}CoO_3$ (Figure 9a) is 0.8 kOe (at 10 K); the low value of $H_C$ signifies the nature of soft ferromagnetic material. Similarly, the ordered $LaBaCo_2O_6$ exhibits a lower value of $H_C$ (0.5kOe) (Figure 9c), whereas the nanoscale-ordered $LaBaCo_2O_6$ phase exhibits a much higher $H_C$ value of 4.2 kOe (Figure 9b) which corresponds to a hard ferromagnet. This different property of the nanoscale-ordered phase was previously explained by a locking of the magnetic spins, due to the presence of 90° oriented nanodomains[16]. The observation of a mottled contrast in the BF images for this form, due to the existence of strain fields, strongly supports this view point. Such strains may induce domain walls[21], which oppose to the reversal of the spins in the applied magnetic field.

Figure 10 shows the temperature dependence of electrical resistivity for the three phases in the presence and absence of an applied magnetic field of ±70 kOe. The resistivity, ρ(T), behavior studied in the 10-400 K range depicts that at high temperature ($T>300K$) the disordered $La_{0.5}Ba_{0.5}CoO_3$ (Figure 10a) and the nanoscale-ordered $LaBaCo_2O_6$ (Figure 10b) phases are semi-metallic, whereas the ordered $LaBaCo_2O_6$ perovskite is clearly semiconducting down to 190 K (Figure 10c). This feature is easily explained by the fact that the first two compounds exhibit 180° Co-O-Co bond angles in this temperature range, in agreement with their cubic or pseudo cubic structural symmetries, favoring a perfect overlapping of the Co 3$d$ orbitals and oxygen 2$p$ orbitals. In contrast to the ordered $LaBaCo_2O_6$ the Co-O-Co bond angles of 174° in the equatorial planes of the $[CoO_2]_\infty$ layers were observed at room temperature[14]. In this case the conduction of charge carriers will be more favorable for linear bond angle, as a result metallic type of conductivity is noticed for the disordered and nanoscale-ordered compounds. With decreasing temperature a transition to a nearly metallic state is observed for the three phases. It is characterized by a change in slope of ρ(T) at $T_C$ for the disordered $La_{0.5}Ba_{0.5}CoO_3$ (Figure 10a) and for the nanoscale-ordered $LaBaCo_2O_6$ (Figure 10b), or by a flat maximum at $T_C$ for the ordered $LaBaCo_2O_6$ (Figure 10c). Thus, these results show that irrespective of their structural nature, the three different forms exhibit a ferromagnetic metallic behavior below $T_C$. Moreover, all the three phases depict an upturn in the resistivity behavior at low temperature. This feature is similar to that observed for the itinerant ferromagnets $SrRuO_3$ and $SrRu_{1-x}M_xO_3$[22,23], which was interpreted as a weak localization contribution associated with electron-electron interaction[24]. In the present case, the magnetoresistance measurements, that will be discussed below, suggest that this upturn is rather due to grain boundary effects. Note that the $\rho(T)$ curves of the disordered $La_{0.5}Ba_{0.5}CoO_3$ and ordered $LaBaCo_2O_6$ are different from those previously reported by Nakajima et al.[14], and corroborate the result reported by Fauth et al.[13] for the disordered $La_{0.5}Ba_{0.5}CoO_3$.

The magnetoresistance (MR) behavior of the three phases (Figure 11) shows a clear magnetic field dependent change in the resistivity below $T_C$. The MR value is calculated as MR (%) = [{$\rho(7)$-$\rho(0)$}/$\rho(0)$]x100, where $\rho(0)$ is the sample resistivity at 0 kOe and $\rho(7)$ under an applied field of ±70 kOe. For the disordered $La_{0.5}Ba_{0.5}CoO_3$ (Figure 11a) and nanoscale-ordered $LaBaCo_2O_6$ (Figure 11b), the maximum MR value is obtained around $T_C$ and the corresponding values are indeed 7 and 6.5% at 179K. The ordered $LaBaCo_2O_6$ exhibits around $T_C$ a rather close value of 6% at 179K (Figure 11c). But importantly, the ordered $LaBaCo_2O_6$ shows an MR value up to 14.5% at 10K in an applied field of ±70 kOe (Figure 11b) which is much larger than an MR value of 4% at the same temperature (10K) for the two other phases (Figure 11a and 11b). Such a difference suggests that at low temperature ($T$<50K), the grain boundary effect plays an important role in the anisotropic MR behavior for the ordered $LaBaCo_2O_6$ perovskite. This is in agreement with its much larger upturn of resistivity, which is almost 10 times larger (at 10K) than the two other phases (Figure 10). Thus, the larger MR observed for the ordered $LaBaCo_2O_6$ can be interpreted as tunnelling magnetoresistance (TMR) effect due to the increase of the intergrain insulating barriers[25], rather than an intrinsic effect, and is dominant at 10K over the TMR effect for the two other phases. Therefore the spin-polarized tunnelling of carriers across the insulating boundaries occurring at the interfaces between polycrystalline grains give rise to the TMR effect in these phases.

Interestingly, the magnetic field dependent isotherm MR behavior at 10 K exhibits an anisotropic effect similar to those of the magnetization behavior (see Figure 9 and 11), which is also present in 50K data. Supporting this behavior, the field dependent MR and M(H) of nanoscale ordered $LaBaCo_2O_6$ is plotted in Figure 12, where the M(H) curve shows a finite coercive field (4.2 kOe) and correlate with the MR data at 10K. The peak in the MR curve occurs at around the coercive field value, which corresponds to the state of maximum domain wall interaction in the orientation of the neighbouring magnetic spins in the 90° oriented domains. The occurrence of anisotropic MR behavior at low temperatures similar to those for M(H) studies suggest the strongly correlated nature of field-induced magnetic and electronic transitions. In contrast to the M(H) loop, the MR loop is strongest for the ordered $LaBaCo_2O_6$ as compared to the other two phases, due to the presence of grain boundaries as discussed earlier. Nevertheless, the isotherm MR data at low temperature for all three phases exhibit hysteresis effects, which resemble the "butterfly-like" feature, although the effect is rather weak for the disordered phase. It is clear from the obtained data that the MR effect is almost isotropic for temperatures near or above the $T_C$ (179, 225 and 300 K). Hence, the butterfly-like feature appears only at low temperatures (studied at 10 and 50 K), which is more prominent in the ordered $LaBaCo_2O_6$. The origin of the magnetic field induced maximum MR near $T_C$ for the disordered $La_{0.5}Ba_{0.5}CoO_3$ and nanoscale-ordered $LaBaCo_2O_6$ phases can be explained by the mechanism of suppression of spin fluctuations below $T_C$. On the other hand, the highest obtained MR at 10 K for the ordered $LaBaCo_2O_6$ is explained by the TMR effect due to the presence of more insulating grain boundaries.

**CONCLUSIONS**

This study shows that the method of synthesis – precursor reactivity and oxygen pressure – plays a crucial role for controlling the cationic order-disorder phenomena in the stoichiometric perovskites $La_{0.5}Ba_{0.5}CoO_3$ (disordered) and $LaBaCo_2O_6$ (ordered and nanoscale-ordered). Importantly, a new form, called nanoscale-ordered $LaBaCo_2O_6$ is obtained in this system, which consists of 112-type 90°-oriented domains. The latter develop strain fields and consequently local lattice distortions which explain the high coercivity of this hard ferromagnet ($H_C \approx 4.2$ kOe) compared to the other two phases which can be considered as soft ferromagnets ($H_C \approx 0.5$ to $0.8$ kOe). From the viewpoint of the transport properties, the disordered La$_{0.5}$Ba$_{0.5}$CoO$_3$ and nanoscale-ordered LaBaCo$_2$O$_6$ exhibit rather similar behavior involving a semi-metal to metal transition around $T_C$, whereas the layer-ordered LaBaCo$_2$O$_6$ is characterized by semi-conductor to metal transition around $T_C$. Moreover, a much larger upturn of the resistivity is observed at low temperature for the latter phase, due to prominent grain boundary effects. As a consequence, the three forms exhibit a similar intrinsic magnetoresistance, maximum at the vicinity of $T_C$ (6 to 7 %), whereas the ordered LaBaCo$_2$O$_6$ exhibits a much larger MR value, indicating tunnelling magnetoresistance at low temperature due to different grain boundary effects.

## ACKNOWLEDGMENT

The authors gratefully acknowledge the CNRS and the Ministry of Education and Research for financial support through their Research, Strategic and Scholarship. E.-L.R. thanks the Finnish Cultural Foundation for financial support.

**TABLES**

**Table 1.** The lattice parameters, calculated cell volumes and selected reliability factors obtained from the XRPD Rietveld refinements.

| $La_{0.5}Ba_{0.5}CoO_3$ | Disordered | Nanoscale Ordered | Ordered |
|---|---|---|---|
| Space Group | $Pm\text{-}3m$ | $Pm\text{-}3m$ | $P4/mmm$ |
| $z$ | 1 | 1 | 2 |
| $a$ (Å) | 3.8863(1) | 3.8855(1) | 3.8997(1) |
| $c$ (Å) | - | - | 7.7158(2) |
| $V$ (Å$^3$) | 58.696(2) | 58.658(2) | 117.291(4) |
| $R_{wp}$ (%) | 4.07 | 4.38 | 2.26 |
| $R_w$Bragg (%) | 6.03 | 5.33 | 5.95 |
| $\chi^2$ | 1.02 | 1.04 | 3.61 |

**Figure 1.** A schematic crystal structure of the 1:1 ordered 112-type LaBaCo$_2$O$_6$ compound.

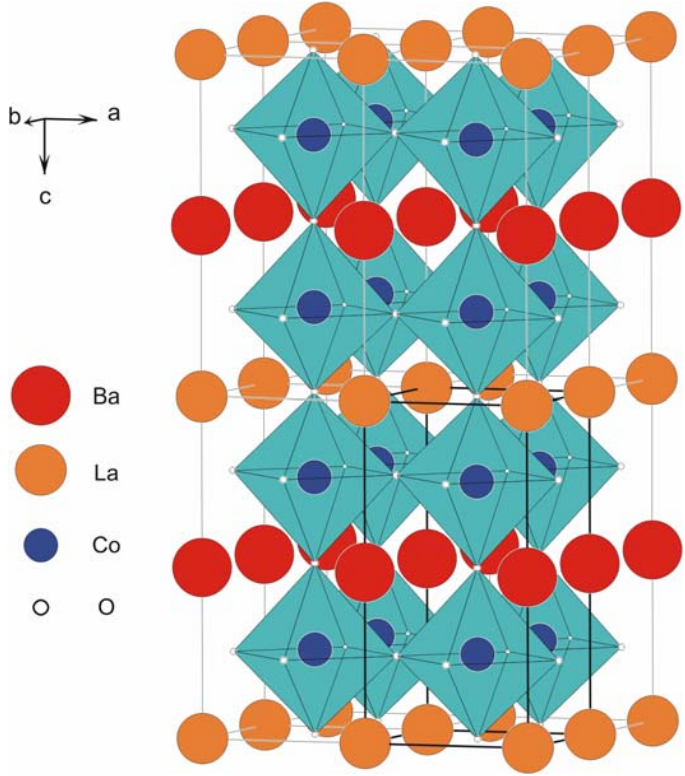

**Figure 2.** The Final observed, calculated and difference plots obtained for the XRPD Rietveld refinement of the (a) disordered (b) nanoscale ordered and (c) ordered LaBaCo$_2$O$_6$ compounds. The inset in each diffractogram shows a magnification of the area where a clear changes in the peak width is observed.

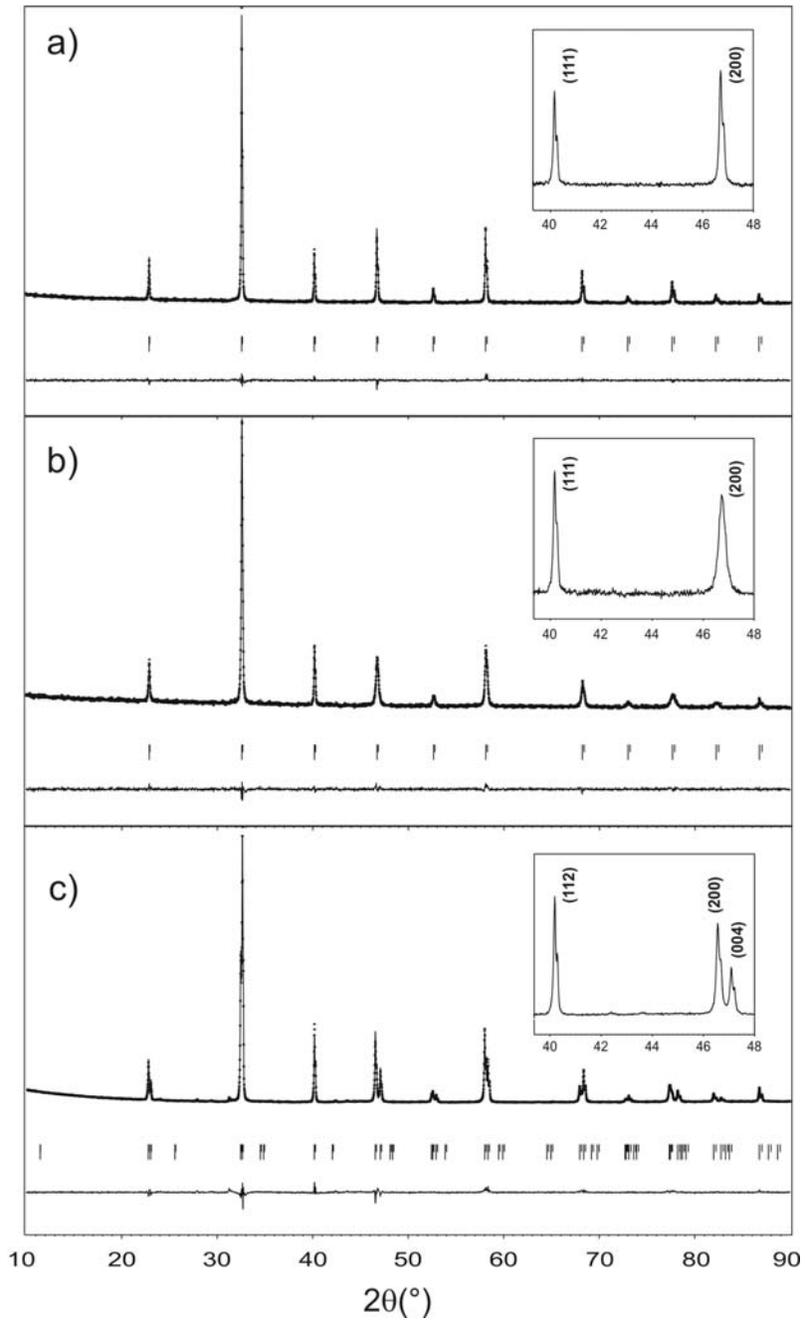

**Figure 3.** Typical HREM images and the corresponding SAED patterns (inset) obtained along the [100] direction respectively for (a) the disordered $La_{0.5}Ba_{0.5}CoO_3$ and (b) the ordered layered $LaBaCo_2O_6$. The 112-type superstructure is clearly identified (b), both in the image and in the SAED patterns, by the doubling of the cell parameter along the [001]* direction as compared to the disordered $La_{0.5}Ba_{0.5}CoO_3$, where only a simple perovskite can be observed along any of the equivalent $<100>_p$ directions.

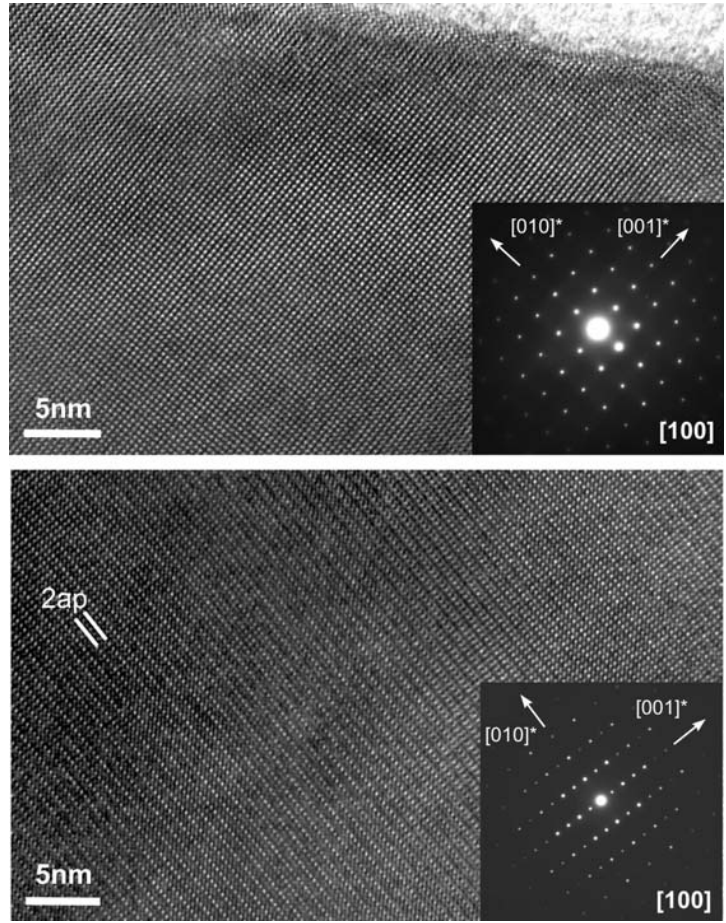

**Figure 4.** Typical SAED patterns observed for the nanoscale-ordered $LaBaCo_2O_6$. The $<100>_p$ zone axis patterns (ZAP) displayed in (a) is the most characteristic. Two sets of supplementary spots compatible with a 112-type supercell are observed. They originate from two 112-type domains (denoted I and II) oriented at 90° the one to the other. A third set of weaker intensity is found at 0 k/2 l/2 positions referring to the perovskite cell. In the $[-210]_I$ ZAP in (b), the rows of weak reflections indicated by arrows are not compatible with the two 112-type domains observed in (a). They are associated with a third 112-type domain (III) whose [001]* direction is oriented at 90° from the [001]* directions of domains I and II. In the $[-110]_I$ in (c), the rows of very weak reflections indicated by arrows are not compatible with any of the three 112-type domains and are associated to a secondary phase denoted X. In (d), a scheme of an archetypal 3D domain texture built up with the three $<100>_p$ 112-type orientation variants is represented.

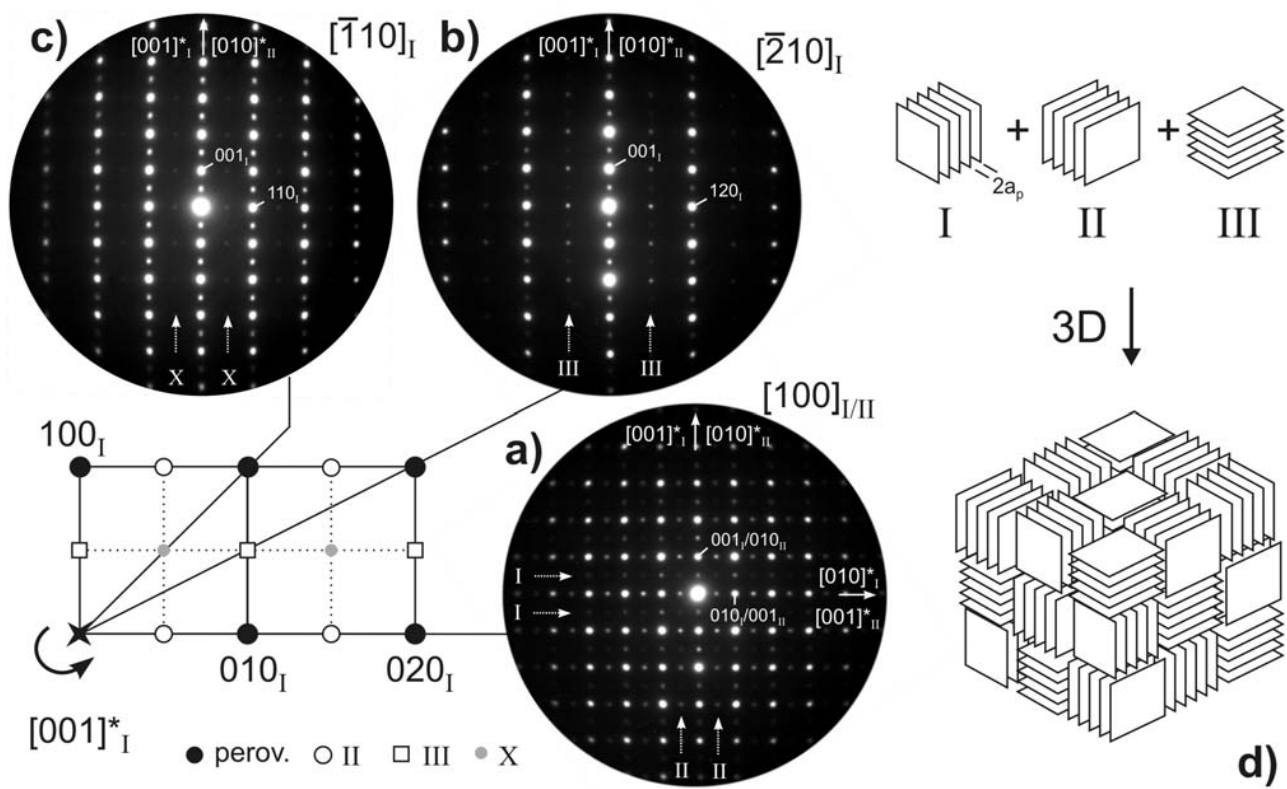

**Figure 5.** Bright-Field images obtained for the tetragonal long-range ordered LaBaCo$_2$O$_6$ (a) and the domain textured nanoscale-ordered LaBaCo$_2$O$_6$ (b). The images are taken in the vicinity of a <110>$_p$ zone axis orientation, tilting around a <100>$_p$* direction and using an objective aperture whose size is chosen to exclude reflections from the perovskite subcell.

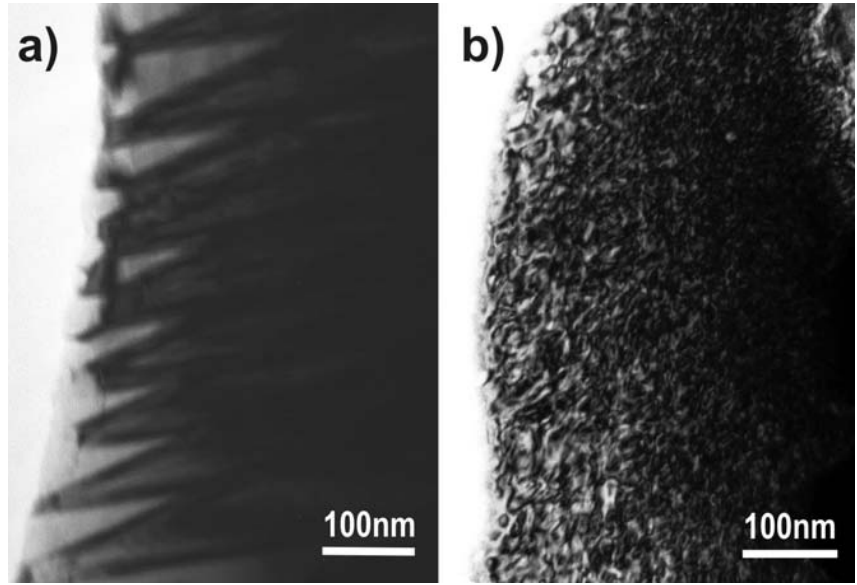

**Figure 6.** Typical Bright Field image (b) observed for the domain textured nanoscale-ordered LaBaCo$_2$O$_6$ compound. This image is obtained in the vicinity of a <100>$_p$ zone axis orientation, tilting around a <110>$_p$* direction and using an objective aperture whose size is chosen to exclude reflections from the perovskite subcell (a). In (c), the enlarged area of (b) allows to evidence the nanometer sized domains having a 2a$_p$ periodicity. By using Bragg filtering in the Fourier space (see the selected Bragg spots in d) one can produce a filtered image in (e) cleaned from the strain field effects and where the domains having a supercell are enhanced. The parts of the image contributing to the formation of the spots at 0 k/2 1/2 positions marked by dashed circles in (e) are indicated by arrows in (d).

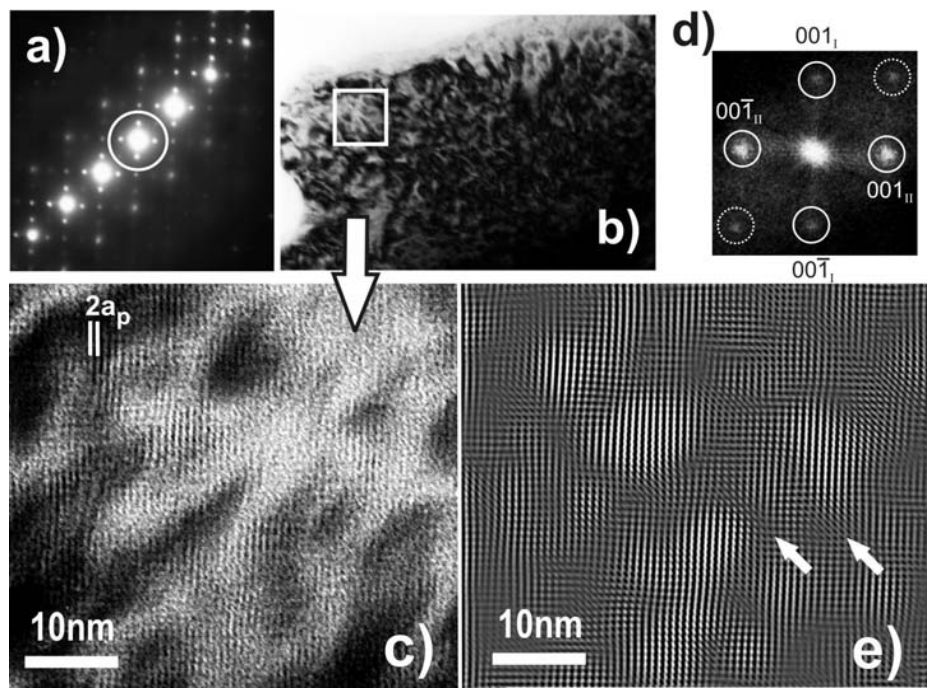

**Figure 7.** (a) <100>$_p$* HREM image showing the 90° oriented domains texture of the nanoscale-ordered LaBaCo$_2$O$_6$. (b) the Fourier transforms obtained from the zones noted 1 to 4 in (a) illustrate how the three orientation variants of 112-type domains can combined to form a 2D domain texture having {100}$_p$ planes as boundary planes. The domain size is typically in the range 5 to 10nm.

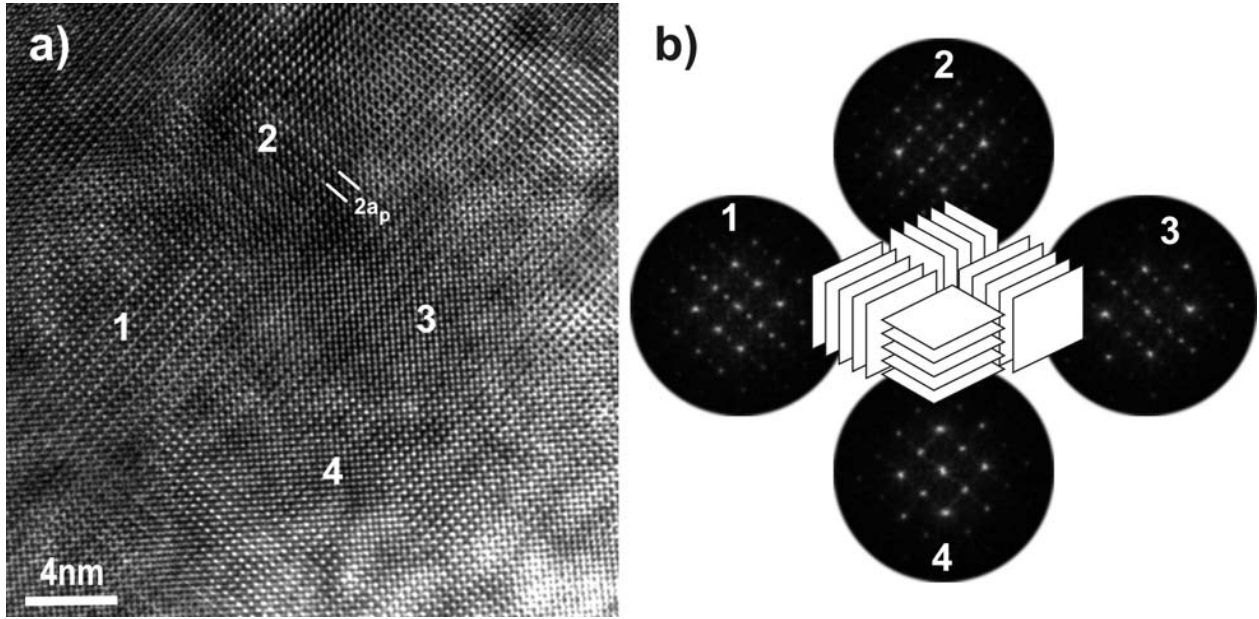

**Figure 8.** Nanoscale-ordered LaBaCo$_2$O$_6$ a) enlarged area of a <100>$_p$* HREM image and the corresponding Bragg filtered image (b). The domain's boundaries are not only formed along {100}$_p$ but also along {110}$_p$, which are usually both combined at a fine scale to produce the 3D domain texture. The image in (b) is produced by selecting in the Fourier space the spots not related to the perovskite subcell.

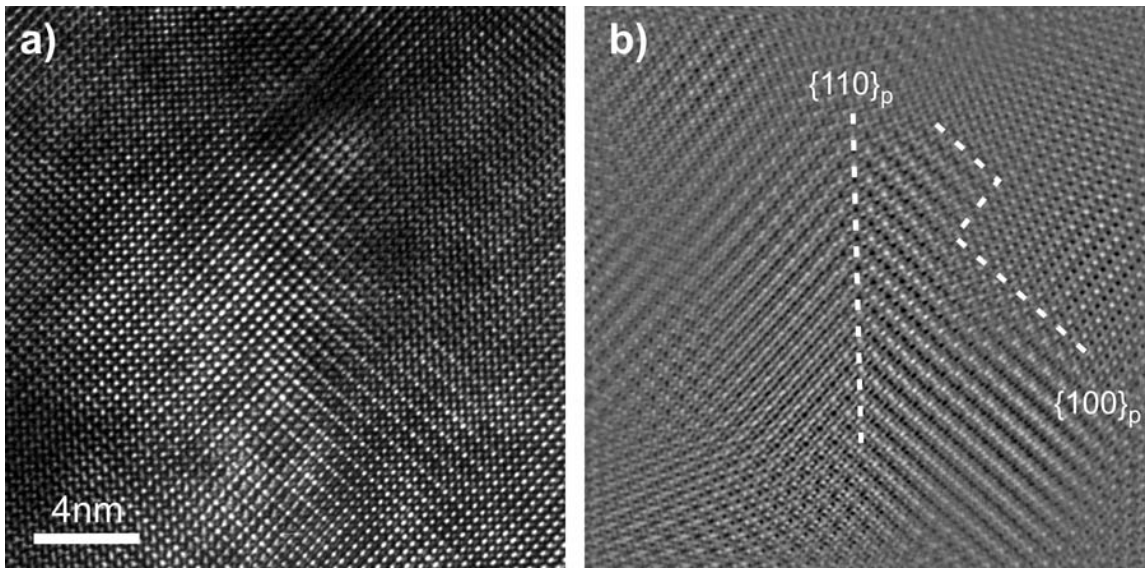

**Figure 9.** Field dependent of isotherm magnetization M(*H*) for (a) disordered La$_{0.5}$Ba$_{0.5}$CoO$_3$ ($\mu_B$ calculated for two f.u.) (b) nanoscale-ordered LaBaCo$_2$O$_6$ and (c) ordered LaBaCo$_2$O$_6$ at different temperatures.

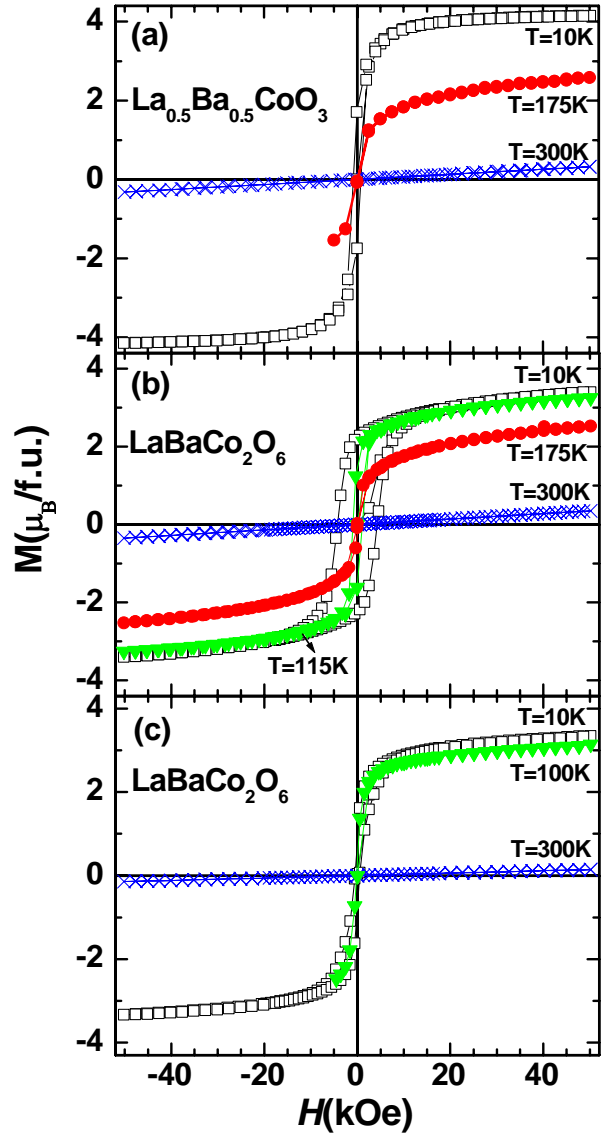

**Figure 10.** Temperature dependent electrical resistivity, ρ, of (a) disordered $La_{0.5}Ba_{0.5}CoO_3$ (b) nanoscale-ordered $LaBaCo_2O_6$ and (c) ordered $LaBaCo_2O_6$ in the presence (solid symbol) and absence (open symbol) of magnetic field (± 70 kOe).

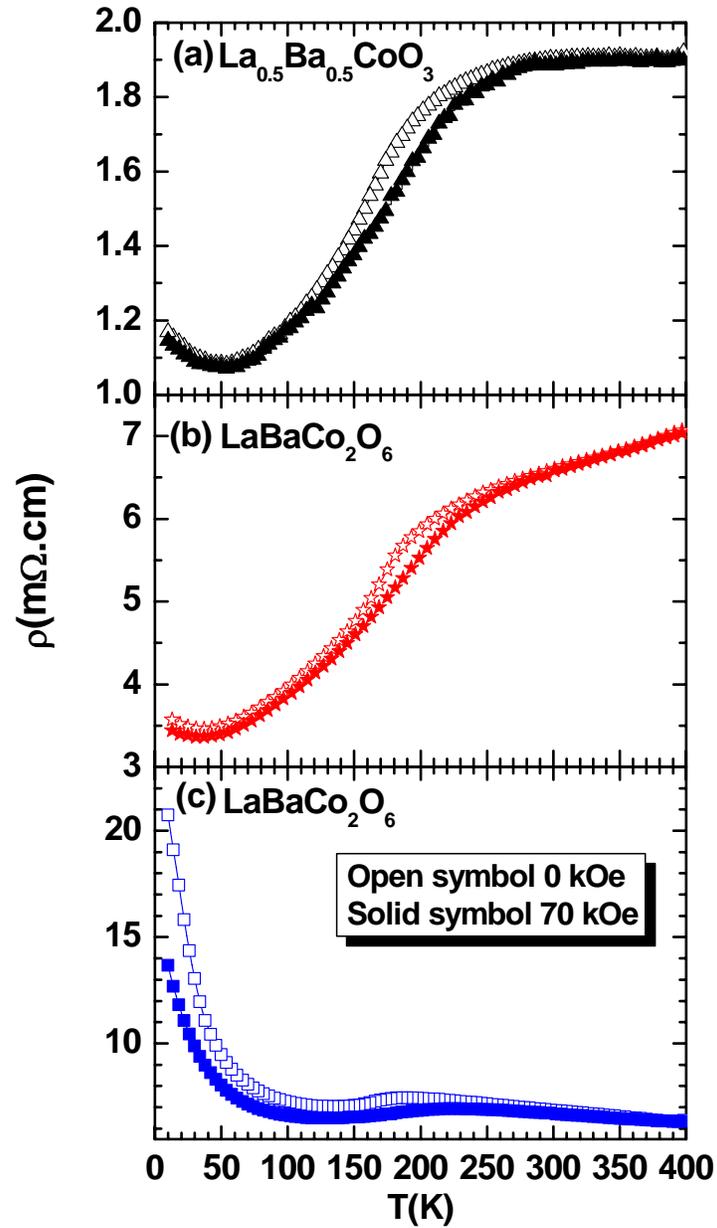

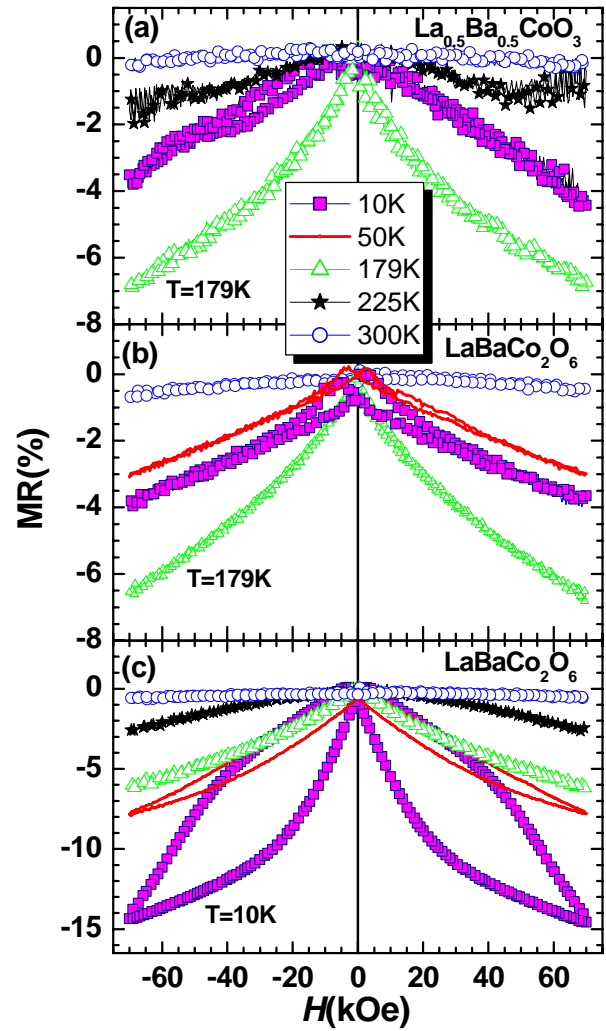

**Figure 11.** Magnetic field dependent isotherm magnetoresistance, MR, effect for (a) disordered $La_{0.5}Ba_{0.5}CoO_3$ (b) nanoscale-ordered $LaBaCo_2O_6$ and (c) ordered $LaBaCo_2O_6$ at different temperatures.

**Figure 12.** Magnetic field dependent isotherm magnetization, M(*H*), and magnetoresistance, MR, for nanoscale-ordered LaBaCo$_2$O$_6$ (at 10K).

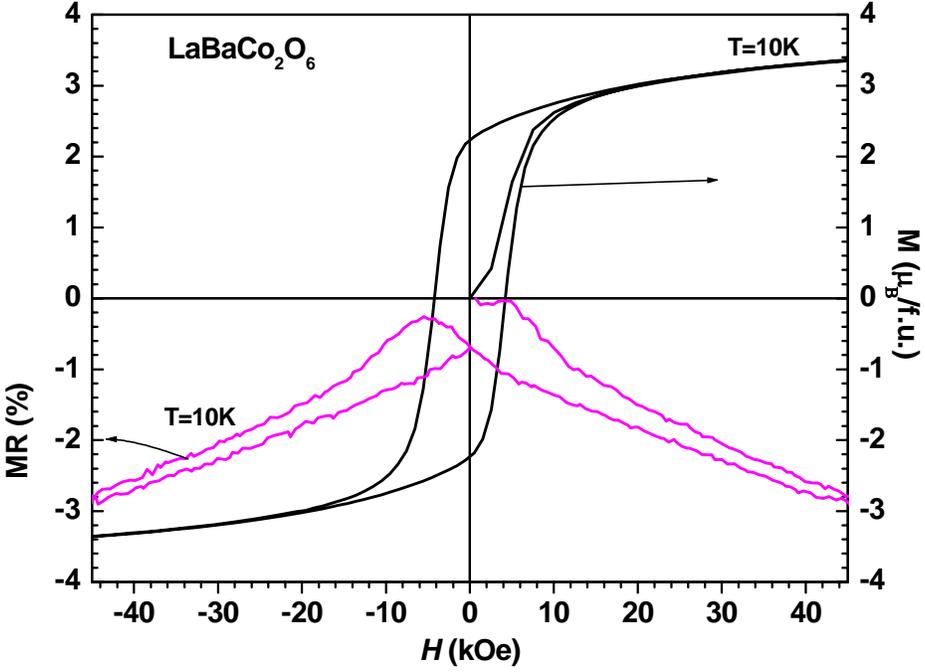